\documentclass{article}
\usepackage[utf8]{inputenc}
\usepackage{amsmath}
\usepackage{color,soul}

\title{Technical Report - Musical Features for Automatic Music Transcription Evaluation}
\author{Adrien Ycart, Lele Liu, Emmanouil Benetos, Marcus T. Pearce}
\date{}

\usepackage[numbers]{natbib}
\usepackage{graphicx}
\usepackage{booktabs}
\usepackage{multirow}
\usepackage{longtable}
\usepackage{url}
\usepackage{hyperref}

\newcommand{\Mod}[1]{\ (\mathrm{mod}\ #1)}

\begin{document}
 
\maketitle

\section{Introduction}

Automatic Music Transcription (AMT) is most often evaluated with metrics that penalise all mistakes equally (see \cite{Bay2009} for a description of common AMT metrics).
However, all mistakes are not equally salient to human listeners: for instance, out-of-key false positives tend to be very noticeable. 
In order to get more insights into the types of mistakes made by a system, we define new, lower-level metrics that can be computed on pairs (target, AMT output).
We aim to define these metrics so that they capture musical aspects that are reported as important by human raters, and mistakes commonly made by AMT systems.

These metrics are defined with polyphonic piano music transcription as the main application.
However, most of them are more general and can be applied in other contexts with minor modifications.

In what follows, we describe the notations we use throughout this document in Section \ref{features:sec:notation}.
% In Section \ref{features:}

\section{Data format}
\label{features:sec:notation}
The AMT output and target are usually represented either as a piano roll or a list of notes.

A piano roll is a $N_p \times T$ binary matrix, where $T$ corresponds to the number of timesteps, and $N_p$ to the number of considered pitches (88 in the case of piano).
We notate the estimated and ground-truth piano rolls as $\hat{M}$ and  $M$ respectively.
Roughly, $M[p,t]=1$ if and only if pitch $p$ is active at timestep $t$.
We notate $M_t$ the binary $N_p$-vector, which corresponds to the $t$-th frame of $M$.

The AMT output and target can also be represented as a list of notes. We notate them $\hat{N}$ and $N$ respectively. They are lists of potentially different lengths $\hat{n}$ and $n$ respectively. Each element of $N$ is a tuple $(s,e,p,v)$ where $s$ and $e$ are the start and end times, $p$ is the MIDI pitch, and $v$ is the original velocity of the note. Each element of $\hat{N}$ is a tuple $(s,e,p)$ (same, without velocity, as it is often not estimated by AMT systems).

\section{Features}

% In this section, we will investigate the various features we can use for that purpose.
% They will be divided into 2 categories: static features (one value for a whole sequence), and dynamic features (a time series).
% We also investigate ways to combine these two types of features.

% We aim at taking (at least) the following things into account:
% \begin{itemize}

% \item Velocity of input notes
% \item Out of key notes (using the same definition of pitch profile)
% \item Octave errors (framewise and notewise)
% \item 19th semitone errors (framewise only)
% \item semitone errors (framewise only)
% \item Repeated and merged notes 
% \item Rhythm mistakes: off-beat rhythm, arpeggiated chords, note duration
% \end{itemize}

% \subsection{Static features}

\subsection{Benchmark Framewise Precision, Recall, F-measure}

These framewise metrics are computed on piano-roll outputs, with a frame duration of 10ms.
We notate for each frame $t$ the number of true positives, false positives and false negatives as respectively, $TP(t)$, $FP(t)$ and $FN(t)$.
A true positive is counted when $M[p,t]=1$ and $\hat{M}[p,t]=1$.
We define the framewise Precision, Recall and F-measure as:

\begin{align}
\mathcal{P} &= \frac{\sum_{t=0}^{T}TP(t)}{\sum_{t=0}^{T}TP(t) + FP(t)} \\[1em]
\mathcal{R} &= \frac{\sum_{t=0}^{T}TP(t)}{\sum_{t=0}^{T}TP(t) + FN(t)} \\[1em]
\mathcal{F} &= \frac{2 \cdot P \cdot R}{P + R} 
\end{align}

\subsection{Benchmark Onset-only Notewise Precision, Recall, F-measure}

These notewise metrics are computed on lists of note outputs.
We count a true positive when a note $(\hat{s},\hat{e},\hat{p})$ is detected and there is a reference note $(s,n,p)$ such that:

\begin{equation}
|\hat{s}-s| < 50ms \land \hat{p} = p
\end{equation}

Each reference note must be matched to at most one estimated note.

We then compute the Precision, Recall and F-Measure as above.
The maximum matching between target and output is computed using \texttt{mir\_eval}\footnote{\url{https://github.com/craffel/mir\_eval}}.

In what follows, we use this definition to determine whether a detected note is a true positive, unless stated otherwise, as it was shown in \cite{TISMIR} that onset-only F-measure is the benchmark evaluation metric that correlates best with human perception (in the case of piano).
When used for other instruments, the onset-offset definition might be preferred.

\subsection{Benchmark Onset-offset Notewise Precision, Recall, F-measure}

These metrics are the same as above, with the difference that a true positive is counted when:

\begin{equation}
|\hat{s}-s| < 50ms \land \hat{p} = p \land |\hat{e}-e|<max(50ms, 0.2*(e-s))
\end{equation}
We then compute Precision, Recall and F-Measure as above.

\subsection{Number of mistakes in highest and lowest voice}
\label{features:sec:highest}

The general idea is that very often, mistakes in the melody or the bassline are more salient than mistakes in middle voices.
We use the highest and lowest voice as a proxy for the melody and the bassline, respectively.
The highest and lowest voices can be defined both framewise or notewise.

Usually, the ground truth is defined so that it corresponds as much as possible to what is contained in the audio signal.
In particular, when the sustain pedal is used, the offsets of notes are usually extended to correspond to the actual duration for which they sound.
When defining the highest and lowest voice, we choose to not take the sustain pedal into account, in order to stick as much as possible to the original partition, and avoid excessive overlapping of notes.
We assume that this is accessible, as most of the time, the sustain pedal is given as an external MIDI control-change parameter, that can either be taken into account or not.

\subsubsection{Framewise}

The highest voice $H_M$ is a time series such that:
\begin{equation}
H_M(t) = \max\limits_{p\in [0,N_p[}({p \mid M[p,t] = 1})
\end{equation}
If $M_t$ is all zeros, we define by default $H_M(t) = -1$.

\begin{itemize}
    \item A true positive is counted for each $(p,t)$ such that $p=H_M(t)$ and $H_M(t)\not=-1$ and $\hat{M}[p,t]=1$.
    \item A false negative is counted for each $(p,t)$ such that $p=H_M(t)$ and $H_M(t)\not=-1$ and $\hat{M}[p,t]=0$.
    \item A false positive is counted for each $(p,t)$ such that $\hat{M}[p,t]=1$ and $p>H_M(t)$. We count all false positives above the highest pitch in the target.
\end{itemize}

With these definitions, we can compute Precision, Recall, and F-measure as before.

The lowest voice can be defined similarly.

\subsubsection{Notewise}

The highest voice $H_N$ is a list of notes such that for all $(s,e,p)$ in $H_N$:
\begin{equation}
\exists t\in [s,e] \mid \forall (s',e',p') \in N, t\not\in [s',e'] \lor p' < p 
\end{equation}

In other words, it is the set of notes that are the highest sounding note at some point in time.
In order to account for cases when, for instance, a chord is slightly arpeggiated and a middle note is played before the highest note, we include a minimum duration $d_H$ for which the note has to be the highest sounding one:

\begin{equation}
\exists t_1,t_2 \in [s,e] \mid t_2-t_1 > d_H \land  \forall (s',e',p') \in N, [t_1,t_2] \cap [s',e'] = \emptyset \lor p' < p 
\end{equation}

\begin{itemize}
    \item A true positive is counted for each note that is a true positive and that is matched to a note in  $H_N$.
    \item A false negative is counted for each note in $H_N$ that is left unmatched.
    
    \item A false positive is counted for each note $(s,e,p)$ in $\hat{N}$ that is left unmatched and such that
    \[
    \exists t\in ]s,e[, \forall (s',e',p') \in N, t\not\in ]s',e'[ \lor p' < p 
    \]
    or with a threshold:
    \[
    \exists t_1,t_2 \in [s,e] \mid t_2-t_1 > d_H \land  \forall (s',e',p') \in N, [t_1,t_2] \cap [s',e'] = \emptyset \lor p' < p 
    \]
    We count all false positives above the highest pitch in the target.

\end{itemize}

We set $d_H=0.5$, this value is set heuristically,  in accordance with the usual threshold for benchmark notewise metrics. 

With these definitions, we can compute Precision, Recall, and F-measure as before.

The lowest voice can be defined similarly.

\subsection{Loudness of false negatives}

The idea is that a missed note will be less noticed if it was played very softly in the input than if it was very loud originally.
It will be even less noticed if there are some louder notes played at the same time.

Similar metrics could be defined with false positives, but since most current AMT systems do not estimate note velocities, we do not take them into account.

We define two such metrics: the normalised false negative loudness, and the false negative loudness ratio.

\subsubsection{Normalised false negative loudness}

For each false negative, we normalise its loudness by the average loudness in a 2-seconds window centered on the onset of the false negative.
In other words, for a given false negative $(s,e,p,v)$ in $N$ define $V$ as:
\begin{equation}
V = \{(s',e',p',v') \mid  |s-s'| < d_L  \}
\end{equation}
where $d_L$ is set as 1 second.
we then compute the normalised loudness as:
\begin{equation}
\text{NormLoud}=\frac{v \times |V|}{\sum_{(s',e',p',v') \in V} v'}
\end{equation}
where $|V|$ denotes the cardinal of $V$.
We then compute the average normalised loudness for all false positives.
It has to be noted that $V$ is never empty; it always contains at least $(s,e,p,v)$.
When $(s,e,p,v)$ is the only note in $V$, the ratio is equal to 1.

\subsubsection{False negative loudness ratio}

We also compute the ratio between the missed note velocity and the loudest note sounding  in the ground truth at the  time of the attack of the missed note.
This ratio is necessarily smaller than 1.
In particular, it is equal to 1 when the missed note is the only one at that time.
We could then average this ratio for all the false negatives.

To do so, we assume an exponential decay of the amplitude of notes, to reflect cases where a loud note is held while other notes are played.
We use an exponential decay rate, applied to the velocities directly.
We stop the decay after 1 second, to avoid notes fading out completely.
Given $a(p)$ the decay rate of a note given its pitch, we define the time varying velocity $v(t)$ of a note (s,e,p,v) as:
\begin{equation}
v(t) = \left\{
    \begin{array}{ll}
        ve^{-a(p)(t-s)} & \mbox{if } t \in [s,s+1] \\
        ve^{-a(p)} & \mbox{if } t \in [s+1,e] \\
        0 & \mbox{otherwise} 
    \end{array}
    \right.
\end{equation}

\begin{figure}[h!]
    \centering
    \includegraphics[width=0.7\columnwidth]{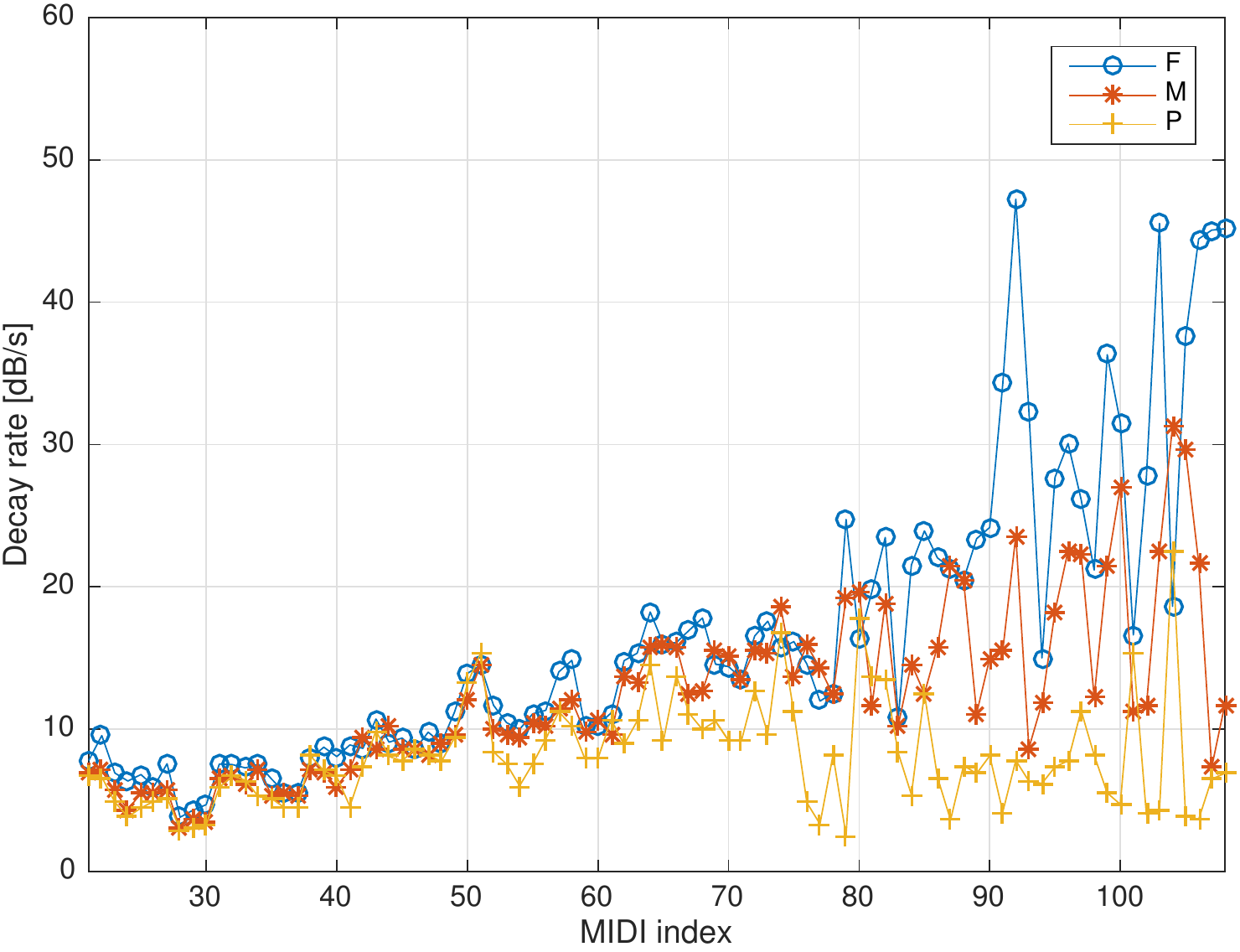}
    \caption{Decay rates per note, for various velocities (taken from \cite{cheng2016exploiting}).}
    \label{features:fig:decayrates}
\end{figure}

The decay rate is obtained from the values presented in Figure \ref{features:fig:decayrates}, taken from  \cite{cheng2016exploiting}.
For each pitch, we average the decay rates across velocities (this has little influence for lower pitches).
Then, we do a linear regression on the averaged decay rates, in order to have smoother, less piano-dependent decay rates estimates.
Using that, we can then compute the velocity of notes through time.
The formula we obtain for $a(p)$ is:
\begin{equation}
a(p) = 0.050532 + 0.021292*p
\end{equation}

The loudness ratio of a false negative $(s,e,p,v)$ is then defined as:
\begin{equation}
\text{LoudRatio} = \frac{v}{
\max\limits_{(s',e',p',v') \in N, t \in [s-d_R,s+d_R]} v'(t) }
\end{equation}
where $d_R$ is a parameter that we set to 50ms.

\subsection{Out-of-key false positives}

The idea is that out-of-key inserted notes are particularly salient.
However, we do not want to rely on key annotations to evaluate that.
Instead, we rely on a pitch profile that is automatically computed from the target.
We propose two versions of this metric, one based on a binary pitch profile, and one on a non-binary pitch profile.

The following definitions assume that the tonality is constant throughout the considered music excerpt.
Behaviour is undefined if that is not the case.

\subsubsection{Binary pitch profile}

We define the binary pitch profile as the set of pitch classes that are active more than 10\% of the time in the target.
This threshold is set heuristically.
Generally-speaking, the higher threshold, the more notes will be considered as out-of-key.
It remains to be seen to what extent varying this threshold influences the metrics.

More precisely, we define an active-pitch-class time series $A_p(t)$ such that:
\begin{equation}
A_p(t) = 1 \Longleftrightarrow \exists (s,e,q)\in N \mid q \equiv p\Mod{12} \land t \in [s,e]
\end{equation}

$A_p(t)$ is defined using the target without pedal, similarly to Section \ref{features:sec:highest}.

We write $P_b$ the set of in-key pitches such that:
\begin{equation}
p \in P_b \Longleftrightarrow  p\equiv q\Mod{12} \land \frac{1}{T} \sum_{t\in [0,T[} A_q(t) > 0.1
\end{equation}

We write $FP_{\hat{N}}$ the total number of false positive notes in $\hat{N}$, $FP_{P_b,\hat{N}}$ the number of out-of-key false positives (i.e. the false positives such that $p\not\in P_b$), and $|\hat{N}|$ the total number of notes in $\hat{N}$.
We then define two ratios:

\begin{equation}
\mathcal{O}_{b,t} = \frac{FP_{P_b ,\hat{N}}}{|\hat{N}|} \textrm{ and } 
\mathcal{O}_{b,p} = \frac{FP_{P_b ,\hat{N}}}{FP_{\hat{N}}}
\end{equation}

$\mathcal{O}_{b,t}$ is the proportion of out-of-key mistakes among all detected notes (and is thus correlated with notewise recall), while $\mathcal{O}_{b,p}$ is the proportion of out-of-key mistakes among all false positives.

\subsubsection{Non-Binary pitch profile}

For each pitch class $q$, we define its fitness to the tonality $F(q)$  as the proportion of the time this pitch class is active:
\begin{equation}
F(q) = \frac{1}{T} \sum_{t\in [0,T[} A_q(t)
\end{equation}

Then, we define the key-disagreement of a false positive $(s,e,p)$ as: $1-F(q)$ for $q$ such that $q \equiv p \Mod{12}$

We then compute 2 values: $\mathcal{O}_{b,t}$ the average key-disagreement of false positives divided by the average key-disagreement of all detected notes, and $\mathcal{O}_{b,p}$ the un-normalised average key-disagreement of false positives.

\subsection{Specific Pitch Errors}\label{features:sec:specific}

A common type of AMT mistake is to have false positives in specific pitch intervals compared to ground-truth notes: semitone errors (neighbouring notes), octave errors (first partial), and 19 semitone errors (second partial).
We define errors for specific pitches $p_e$ with $p_e\in \{1,12,19\}$.
We define these metrics both framewise and notewise.

\subsubsection{Framewise}

$\hat{M}[p,t]$ is a specific pitch error if and only if the following conditions are all true:

\begin{itemize}
    \item $\hat{M}[p,t]= 1$
    \item $M[p,t]= 0 $
    \item $M[p-p_e,t]= 1 \vee M[p+p_e,t]= 1$
    \item $\forall i \in [t-\delta, t[, M[p,i] = 0$
\end{itemize}
where $\delta$ is a parameter that we set heuristically to 50ms, in accordance with the threshold used for benchmark notewise metrics.
The last condition is to ensure that erroneous continuations of correct notes are not penalised if there was also a target note $p_e$ semitones apart right after.

For $p_e=19$, we only consider ground-truth notes 19 semitones below, as second partial mistakes usually only happen above the ground truth notes. The third condition becomes simply: 
$ M[p-p_e,t]= 1 $

We then compute two different ratios: let $N_s$ be the number of specific pitch errors, $N_f$ the number of frames, and $N_e$ the total number of false positives:
\begin{equation}
\mathcal{S}_{p_e,f,t} = \frac{N_s}{N_f} \textrm{ and } \mathcal{S}_{p_e,f,p} = \frac{N_s}{N_e}
\end{equation}

The first is correlated to $\mathcal{P}$, while the second can be biased when $\mathcal{P}$ is very high (an output with only 1 error that happens to be a specific pitch error will have $\mathcal{S}_{f,p}=1$)

\subsubsection{Notewise}

A note $(s,e,p)$ is a specific pitch false positive if:
\begin{itemize}
    \item $(s,e,p)$ is a false positive
    \item $\exists (s',e',p') \in N \mid |p-p'|=p_e \land 
\frac{\min(e,e') - max(s,s')}{e-s} > 0.8$
\end{itemize}

The last condition boils down to saying that there is a ground truth note $p_e$ semitones apart from $(s,e,p)$ that overlaps with $(s,e,p)$ for 80\% of $(s,e,p)$'s duration.
The value 80\% was set heuristically, and echoes the benchmark onset-offset notewise metrics condition requiring that the offset of a note is within 20\% of its duration.

For $p_e=19$, we only consider ground truth notes 19 semitones below $(s,e,p)$, for the same reason as above.

Similarly as framewise metrics, we compute 2 ratios: the proportion of specific pitch mistakes among all detected notes, and among false positives.

\subsection{Repeated and merged notes}

Another common type of mistake in AMT is to have repeated (i.e. fragmented) notes, or incorrectly merged notes.

\subsubsection{Repeated notes}

A note $(s,e,p) \in \hat{N}$ is counted as a repeated note when it fulfills the following conditions:

\begin{itemize}
    \item It is an Onset-only false positive
    \item $\exists (s',e',p') \in N \mid  p=p' \land \frac{\min(e,e') - max(s,s')}{e-s} > 0.8$
    \item $\exists (s'',e'',p'') \in \hat{N}$ such that:
        \begin{itemize}
            \item  $(s'',e'',p'') \not= (s,e,p)$
            \item $p''=p'$
            \item $\frac{\min(e'',e') - max(s'',s')}{e''-s''} > 0.8$
            \item $e''<s$
        \end{itemize}
    
\end{itemize}

Put more simply, a note is considered as a repeated note if it is a false positive, if it overlaps with a ground-truth note, and if there is another previous detected note that overlaps with the same ground-truth note.

We can then compute either the proportion of repeated notes among false positives, or among all notes, as with the previous metrics.
Once again, for both of these metrics, the threshold 80\% was set heuristically, and echoes the benchmark onset-offset notewise metrics condition requiring that the offset of a note is within 20\% of its duration.

\subsubsection{Merged notes}

A note $(s,e,p) \in N$ is counted as a merged note when it fulfills the following conditions:

\begin{itemize}
    \item It is an Onset-only false negative
    \item $\exists (s',e',p') \in \hat{N} \mid  p=p' \land \frac{\min(e,e') - max(s,s')}{e-s} > 0.8$
    \item $\exists (s'',e'',p'') \in N$ such that:
    \begin{itemize}
        \item  $(s'',e'',p'') \not= (s,e,p)$
        \item $p''=p'$
        \item $\frac{\min(e'',e') - max(s'',s')}{e''-s''} > 0.8$
        \item $e''<s$
    \end{itemize}
\end{itemize}

A note is thus considered as a merged note if it is a false negative, if it overlaps with a detected note, and if there is another previous ground-truth note that overlaps with the same detected note.

We can then compute either the proportion of merged notes among false negatives, or among all notes, as with the previous metrics.

\subsection{Rhythm features}

\subsubsection{Rhythm histogram spectral flatness}
\label{features:sec:rhythm_hist}

Rhythm is an important aspect of music.
We thus define a metric to account for rhythmic imprecision as follows.
We define $O$ and $\hat{O}$ the (ordered) list of note onsets of $N$ and $\hat{N}$ respectively, potentially with repetition.
$O$ is of same size as $N$.

Based on these, we compute inter-onset-intervals ($IOI$ and $\hat{IOI}$) as the first derivative of $O$ and $\hat{O}$ respectively. Let $n$ be the number of notes:
\begin{equation}
\forall 0\leq i<n-1,  IOI(i)= O(i+1)-O(i)
\end{equation}

We then compute a normalised histogram of the IOIs, with bins as follows: from 0 to 100ms, we use a bin size of 10ms, and from 100ms to 2s, we use a bin size of 100ms.
Overall, we have 29 bins.
We notate this list $b$, and the resulting histogram $h_r$.
The spacing in bins is set heuristically; however, it could have a strong influence on the result.
We leave it to future work to quantify that influence.

From this IOI, we compute its spectral flatness \cite{johnston1988transform}, which is defined as the ratio of the geometric mean of the histogram over its arithmetic mean:
It is used usually on power spectra, and represents the peakiness of the spectrum.
It is useful in our case, as quantised rhythms would give an IOI histogram with only some non-zero bins, corresponding to specific note values, while rhythm imprecision would spread the values across several bins.
We thus have:
\begin{equation}
S_f = \frac{1}{29}\sum_{0\leq i<29}\log(h_r[i]) - \log\biggl(\frac{1}{29}\sum_{0\leq i<29}h_r[i]\biggr)
\end{equation}

In practice, we add $\epsilon = 10^-5$ to deal with 0 values in $h_r$.

We then compute the spectral flatness value for $N$ and $\hat{N}$.
We use as features the spectral flatness for $\hat{N}$ and the difference between the spectral flatness of $\hat{N}$ and $N$.

\subsubsection{Rhythm dispersion}

Another approach to attempt to characterise rhythmic deviations is to run K-means clustering \cite{murphy2012machine} on the IOI set.
Ideally, each cluster would correspond to one note value, with small variations due to tempo deviations and interpretation mostly.
We can then assess the mean and standard deviation within each cluster.

Setting the number of clusters here is a tough problem.
If we do not have enough clusters, one cluster might correspond to several note values, which would result in artificially high standard deviation.
If we have too many clusters, we might end up with several clusters corresponding to the same note value, or in the extreme case, one cluster per note.

To determine the number of clusters and their initial centers, we compute a normalised IOI histogram on the target, similar to the previous, but with higher bin size: 20ms between 0 and 0.1s, and 200ms between 0.1 and 2s.
We then choose as initial cluster centres all the peaks in the resulting $h_r$.
Here again, the spacing in bins is set heuristically; however, it has to be noted that it might have a strong influence on the number of clusters.

We first run K-means clustering on the target $IOI$ set.
After convergence, we use the resulting cluster means as initial values to run K-means clustering on the estimated $IOI$ set.
We then compute the distance between cluster means for the estimated and target $IOI$ sets, and the relative difference between the cluster standard deviations for the estimated and target $IOI$ sets.
We use as feature the mean, maximum and minimum across clusters, for both the centre drifts and standard deviation differences.

\subsubsection{Validating Rhythm features}

We have seen in the experiments presented in \cite{TISMIR} that these rhythmic features have a high importance when modelling perceptual ratings of AMT quality.
In order to validate that these metrics do capture rhythm deviations, we run some experiments.

We use as target the AMT outputs for all the stimuli in presented in \cite{TISMIR}.
We use as outputs various modified versions of these same MIDI files, by order of rhythm regularity (high to low):

\begin{description}
    \item[Quant-constant:] Quantised MIDI files with 16th note precision, using a constant tempo equal to the average tempo over the whole segment (we use the A-MAPS tempo and beat annotations described in \cite{ycart2018maps});
    \item[Quant:] Quantised MIDI files with 16th note precision, using a time-varying tempo, with the ground-truth 16th note positions;
    \item[Noisy-100:] Add uniform noise in [-100ms,100ms] to the onset times;
    \item[Noisy-300:] Add uniform noise in [-300ms,300ms] to the onset times.
\end{description}

We report in Table \ref{features:tab:rhythm_results} the mean and standard deviation (std) of the rhythm features in each condition.
It appears that the features behave generally as expected.
The mean spectral flatness is lowest for Quant-constant, and highest for the Noisy configurations.
Besides, the spectral flatness difference is negative for the quantised versions, and positive for the noisy versions.
For the rhythm dispersion values, we see a similar trend: for quantised versions, the average change in std is negative, while it is positive for noisy versions.
Moreover, the greater the noise, the greater the average change in std.

However, it appears that increasing the noise level does not change the mean spectral flatness of the outputs, which is kind of surprising.
This might be due to the bin size we used: since we use bins of size 100ms between 100ms and 2s, small differences in noise might be hard to catch.
Another possibility is that since we use short examples, in a lot of cases, histogram bins contain one single value.
Adding noise to that value will change the bin it is counted in, but will not change the overall spectral flatness.
This might also explain why the dispersion average drift also increases with the noise level: although the noise is centred on zero, it is likely applied to many clusters with one single, or few values in it, so the drift does not always cancel out on average within a cluster.

It also appears that the dispersion feature values are very similar for both quantised versions. This might be due to the fact that we use short segments, and that tempo variations are quite small usually, so there is probably little difference between the two quantised conditions.

\begin{table}
 \begin{center}
%  \hspace*{-2cm}
\resizebox{\textwidth}{!}{
 \begin{tabular}{ r || c c | c c | c c | c c }
 \toprule
 \multirow{2}{*}{\bfseries Feature} & \multicolumn{2}{c}{\bfseries Quant-constant}& \multicolumn{2}{|c}{\bfseries Quant} & \multicolumn{2}{|c}{\bfseries Noisy-100} & \multicolumn{2}{|c}{\bfseries Noisy-300} \\
& mean & std & mean & std& mean & std& mean & std\\
\midrule
\midrule
Spectral Flatness Output & \textit{-9.842} & 0.857 & -9.572 & 0.909 & \textbf{-6.695 }& 0.814 & -6.751 & 0.772 \\
Spectral Flatness Difference& \textit{-1.850} & 0.973 & -1.580 & 0.898 & \textbf{1.298} & 1.217  & 1.241 & 1.380 \\
\midrule
Dispersion Avg. std Change & \textit{-0.010} & 0.019 & \textit{-0.010} & 0.017 & 0.024 & 0.020  & \textbf{0.080} & 0.038 \\
Dispersion Min. std Change &\textit{ -0.032} & 0.036 & -0.029 & 0.041 & -0.001 & 0.050  & \textbf{0.037} & 0.074 \\
Dispersion Max. std Change & 0.012 & 0.031 & \textit{0.009} & 0.026 & 0.049 & 0.027 & \textbf{0.125} & 0.048 \\
Dispersion Avg. Drift      & \textit{0.025} & 0.022 & \textit{0.025} & 0.026 & 0.038 & 0.032  & \textbf{0.129} & 0.047 \\
Dispersion Min. Drift    & \textit{0.008} & 0.009 & \textit{0.008} & 0.009 & 0.015 & 0.013  & \textbf{0.057} & 0.045 \\
Dispersion Max. Drift     & \textit{0.046} & 0.050 & \textit{0.046} & 0.059 & 0.065 & 0.068 & \textbf{0.209} & 0.095 \\

\bottomrule
 \end{tabular}}
\end{center}
 \caption{Feature means and standard deviation (std) across all stimuli, with 4 levels of rhythmic precision. Highest mean values are in bold, lowest mean values are in italic.}
 \label{features:tab:rhythm_results}
% \vspace{-0.3cm}
\end{table}

\subsection{Consonance measures}

We choose 3 different consonance measures: one based on periodicity/inharmonicity, one based on partials interference, and one based on culture (statistical frequency in a corpus).
These are computed using Peter Harrison's implementation\footnote{https://github.com/pmcharrison/incon}.
In particular, we use the following features:
\begin{itemize}
    \item \texttt{hutch\_78\_roughness} for partials interference
    \item \texttt{har\_18\_harmonicity} for periodicity
    \item \texttt{har\_19\_corpus} for culture.
\end{itemize}
These 3 consonance measures were shown to correlate best with perceptual ratings of consonance \cite{harrison_pearce_2019}.

We then compute these consonance measures on the output and target piano rolls, using an \emph{event} timestep: 
one timestep per new onset or offset.
The above features are undefined for silence, we thus do not take them into account in the computations.
We then compute the weighted average (using as weight each frame's duration in sections), the weighted standard deviation, minimum and maximum value for each feature, both on the output and target piano rolls.
We use as features the weighted average, the weighted standard deviation, minimum and maximum computed on each consonance measures on the output piano roll.

\subsection{Polyphony level}

We assume that a mistake is more salient when it is the only note being played.
Conversely, if a big chord is supposed to be played, but few notes are detected, this will be noticeable.

We compute the difference in polyphony level as a time series:

\begin{equation}
\text{Poly}(t) = \biggl|\sum_{0\leq p<88} \hat{M}[p,t]-\sum_{0\leq p<88} M[p,t]\biggl|
\end{equation}

We then use as features the mean, standard deviation, minimum and maximum of this series.

\section{Summary} 
 
 We provide a table summarising all the features.
 The first column corresponds to feature groups (as described in the sections above), the second column describes each scalar value that can be found within that feature group, and the last column describes whether higher is better for that metrics: ``Yes'' if higher is better, ``No'' if lower is better, ``/'' when it depends on other factors.
 
% \begin{table}[h]
% \centering
  \begin{longtable}{p{3cm}|p{5cm}|c}
  \toprule
  \bfseries Feature  group & \bfseries Sub-feature & \bfseries  Higher is better? \\
\midrule
\endhead
\multirow[t]{3}{=}{Benchmark Framewise metrics} 
 & Precision  & Yes \\
% \cline{2-4}
&  Recall  & Yes \\
% \cline{2-4}
& F-measure  & Yes \\
\hline
\multirow[t]{3}{=}{Benchmark Onset-only notewise metrics} & Precision  & Yes \\
% \cline{2-4}
&  Recall  & Yes \\
% \cline{2-4}
& F-measure  & Yes \\
\hline
\multirow[t]{3}{=}{Benchmark Onset-Offset notewise metrics} & Precision  & Yes \\
% \cline{2-4}
&  Recall  & Yes \\
% \cline{2-4}
& F-measure  & Yes \\
% \hline
\midrule
\multirow[t]{3}{=}{Framewise mistakes in highest voice} & Precision  & Yes \\
% \cline{2-4}
&  Recall  & Yes \\
% \cline{2-4}
& F-measure  & Yes \\
\hline
\multirow[t]{3}{=}{Framewise mistakes in lowest voice} & Precision  & Yes \\
% \cline{2-4}
&  Recall  & Yes \\
% \cline{2-4}
& F-measure  & Yes \\
\hline
\multirow[t]{3}{=}{Notewise mistakes in highest voice} & Precision  & Yes \\
% \cline{2-4}
&  Recall  & Yes \\
% \cline{2-4}
& F-measure  & Yes \\
\hline
\multirow[t]{3}{=}{Notewise mistakes in lowest voice} & Precision  & Yes \\
% \cline{2-4}
&  Recall  & Yes \\
% \cline{2-4}
& F-measure  & Yes \\
\midrule

\multirow[t]{2}{=}{Loudness} & Normalised false negative loudness & No \\
& False negatives loudness ratio  & No \\

\midrule

\multirow[t]{2}{=}{Binary out-of-key false positives} & Proportion among false positives  & No \\
& Proportion among detected notes  & No \\
\hline
\multirow[t]{2}{=}{Non-binary out-of-key false positives} & Average key-disagreement of false positives  & No \\
& Average key-disagreement of false positives normalised by the average key-disagreement of all detected notes & No \\

\midrule
\multirow[t]{2}{=}{Framewise semitone errors} & Proportion among false positives  & / \\
& Proportion among detected notes  & / \\
\hline
\multirow[t]{2}{=}{Framewise octave errors} & Proportion among false positives  & / \\
& Proportion among detected notes  & / \\
\hline
\multirow[t]{2}{=}{Framewise third-harmonic errors} & Proportion among false positives  & / \\
& Proportion among detected notes  & / \\
\hline
\multirow[t]{2}{=}{Notewise semitone errors} & Proportion among false positives  & / \\
& Proportion among detected notes  & / \\
\hline
\multirow[t]{2}{=}{Notewise octave errors} & Proportion among false positives  & / \\
& Proportion among detected notes  & / \\
\hline
\multirow[t]{2}{=}{Notewise third-harmonic errors} & Proportion among false positives  & / \\
& Proportion among detected notes  & / \\

\midrule

Repeated notes & Proportion among false positives 
& / \\
 & Proportion among detected notes  & / \\
 \hline
 Merged notes & Proportion among false positives 
& / \\
 & Proportion among detected notes  & / \\

\midrule
\multirow[t]{2}{=}{Rhythm histogram spectral flatness} & Value computed on output & / \\
 & Relative difference between value computed on output and on target & / \\
\hline
\multirow[t]{2}{=}{Rhythm dispersion}& Mean centre drift & No \\
 & Minimum centre drift & No \\
 & Maximum centre drift & No \\
 & Mean cluster standard deviation difference & / \\
 & Minimum cluster standard deviation  difference & / \\
 & Maximum cluster standard deviation  difference & / \\
 
\midrule
 \multirow[t]{2}{=}{Consonance measures}& Mean of \texttt{hutch\_78\_roughness} & / \\
 & Standard deviation of \newline \texttt{hutch\_78\_roughness} & / \\
& Minimum of \newline \texttt{hutch\_78\_roughness} & / \\
& Maximum of \newline \texttt{hutch\_78\_roughness} & / \\
\cline{2-3}
& Mean of \texttt{har\_18\_harmonicity} & / \\
 & Standard deviation of \newline \texttt{har\_18\_harmonicity} & / \\
& Minimum of \newline \texttt{har\_18\_harmonicity} & / \\
& Maximum of \newline \texttt{har\_18\_harmonicity} & / \\
\cline{2-3}
& Mean of \texttt{har\_19\_corpus} & / \\
 & Standard deviation of \newline \texttt{har\_19\_corpus} & / \\
& Minimum of \texttt{har\_19\_corpus} & / \\
& Maximum of \texttt{har\_19\_corpus} & / \\

  \bottomrule

       \caption{
       Summary of all the proposed evaluation metrics.}

  \end{longtable}

% \end{table}

% \todo{MAKE A TABLE THAT RECAPS THE METRICS, THEIR NAMES (maybe), THEIR VALUES, AND WHETHER HIGHER IS BETTER (if time allows...)}

\bibliographystyle{plain}
\bibliography{main}
\end{document}